# Generating Subsurface Earth Models using Discrete Representation Learning and Deep Autoregressive Network

Jungang Chen

*Harold Vance Department of Petroleum Engineering, College of Engineering, Texas A&M University, College Station, Texas, USA*

Chung-Kan Huang and Jose F. Delgado

*ConocoPhillips, Houston, Texas, USA*

and

Siddharth Misra, Ph.D.

misra@tamu.edu

*Associate Professor, Harold Vance Department of Petroleum Engineering, College of Engineering, Texas A&M University, College Station, Texas, USA*

*Associate Professor (joint appointment), Department of Geology and Geophysics, College of Geosciences, Texas A&M University, College Station, Texas, USA*

# Generating Subsurface Earth Models using Discrete Representation Learning and Deep Autoregressive Network

**ABSTRACT**

Subsurface earth models (referred as geomodels) are crucial for characterizing complex subsurface systems. Multiple-point statistics is commonly used to generate geomodels. In this paper, a deep-learning-based generative method is developed as an alternative to traditional geomodel generation procedure. The generative method comprises two deep-learning models, namely hierarchical vector-quantized variational autoencoder (VQ-VAE-2) and PixelSNAIL autoregressive model. Based on the principle of neural discrete representation learning, the VQ-VAE-2 learns to massively compress the geomodels to extract the low-dimensional, discrete latent representation corresponding to each geomodel. Following that, PixelSNAIL uses deep autoregressive network to learn the prior distribution of the latent codes. For the purpose of geomodel generation, PixelSNAIL samples from the newly learnt prior distribution of latent codes, and then the decoder of the VQ-VAE-2 converts the newly sampled latent code to a newly constructed geomodel. PixelSNAIL can be used for unconditional or conditional geomodel generation. In unconditional generation, the generative workflow generates an ensemble of geomodels without any constraint. On the other hand, in the conditional geomodel generation, the generative workflow generates an ensemble of geomodels similar to a user-defined source image, which ultimately facilitates the control and manipulation of the generated geomodels. To better construct the fluvial channels in the geomodels, perceptual loss is implemented in the VQ-VAE-2 model instead of the traditional 'mean squared error' loss. At a specific compression ratio, the quality of multi-attribute geomodel generation is better than that of single-attribute geomodel generation.

## 1        REVIEW OF GEOMODEL COMPRESSION AND GENERATION TECHNIQUES

A geomodel is a large-scale digital representation of the subsurface earth. Geomodelling is commonly used for managing natural resources, identifying natural hazards, and quantifying geological processes. Geomodels are commonly used to better characterize and plan the development of oil and gas fields, groundwater aquifers and ore deposits. Geomodels of subsurface reservoir properties, such as porosity, permeability, and saturation, are stochastically created by combining static information from sources (core samples, seismic data, well logs, and geologic records) with dynamic data from production and well tests. Some subsurface properties, like porosity and permeability, are assigned to each individual grid block, while others, like relative permeability and capillary pressure, are defined either by layer or for the entire model.

Because subsurface earth has a vast spatial area, geomodels are typically large in size and contain millions of cells in order to accurately represent the complexities and heterogeneity present in the subsurface. This results in higher computational costs both for generating the geomodel and for using it for forecasting. After being generated, the geomodels are adjusted to match observed field data through a process known as history matching. However, the significant size of these models greatly increases the computational cost of history matching and forecasting. In the oil and gas industry, realistic geomodels are required as input to reservoir simulator programs, which predict the reservoir behavior under various hydrocarbon recovery scenarios. The use of geomodels and reservoir simulation can aid in determining the best recovery option for a reservoir in terms of safety, efficiency, effectiveness, and economics. These models help identify a development plan that is optimized for a particular reservoir.

## 1.1 Geomodel Compression using Neural Networks

In order to reduce computational cost when using geomodels with millions of computation grid blocks, low-dimensional representations of these geomodels are computed as an alternate or proxy model by using dimensionality reduction techniques. Principal component analysis (PCA) is a widely used dimensionality reduction technique for large datasets, it projects high manifold data into lower dimensional linear space by extracting the directions that explain the maximum variance of data [1]. Singular value decomposition (SVD) is another popular dimensionality reduction method. SVD factorizes real/complex matrix into left and right unitary matrices and a rectangular diagonal matrix containing non-negative values, also referred as singular values. SVD is used for dimensionality reduction by selecting certain number of singular values to approximate the original matrix. Singular values computed using SVD are analogous to the principal components computed using PCA. Neural networks are also used for dimensionality reduction. Authors used traditional autoencoders (AE), such as variational autoencoder (VAE), and advanced autoencoders, such as vector-quantized variational autoencoder (VQ-VAE) and hierarchical VQ-VAE, referred as VQ-VAE-2, to find low dimensional representation of geomodels [2, 3].

### 1.1.1 Autoencoder (AE)

Autoencoders are unsupervised neural networks that can learn the nonlinear, low-dimensional representations of large spatial or sequential data, such as images. Autoencoders have been applied to a wide range of tasks, including dimensionality reduction, image denoising, generative models, and anomaly detection, among others. In an autoencoder, the encoder and decoder can be thought of as two separate neural networks that work in tandem. The encoder compresses the input data into a lower-dimensional representation, while the decoder expands the lower-dimensional representation back to the original data. The encoding and decoding processes are implemented as different neural network layers, where the intermediate layer between the encoder and decoder is referred to as the bottleneck or latent space [4]. The autoencoder is trained by minimizing a reconstruction loss, which measures the difference between the output of the decoder and the original input data. The goal of the training process is to learn a compact and meaningful representation of the data in the bottleneck layer.

### 1.1.2 Variational Autoencoder (VAE)

The Variational Autoencoder (VAE) is similar to Autoencoder (AE) and also comprises of an encoder network and decoder network. The encoder maps the input data to a multivariate latent distribution instead of a single latent vector, and the decoder samples a latent vector from the latent distribution and tries to reconstruct the input data as accurately as possible [5]. The latent space of a VAE is regularized by adding a Kullback-Leibler (KL) divergence to the training loss. The multivariate normal distribution of its latent vector makes it continuous, centered, and easier to interpolate compared to the latent vector of an AE. A VAE learns a latent distribution that is Gaussian, and thus it can generate new data by sampling a new latent vector. In other words, the decoder of a VAE can be used to generate new samples not present in the training dataset because the VAE latent space is continuous, centered, and evenly spread.

### 1.1.3 Vector-Quantized Variational Autoencoder (VQ-VAE)

VQ-VAE is different from AE and VAE in that it uses a discrete latent space for representation learning [6]. This is more suitable for sparse data, such as images, audio, and videos that can be represented as combinations of discrete information, such as discrete objects in images and specific sounds in audio. Unlike VAE, VQ-VAE alone does not have the capability of reliable generation, so it needs to be combined with autoregressive methods such as PixelCNN or PixelSNAIL to achieve this. This generation approach eliminates the issue of posterior collapse in VAE, which can lead to blurry reconstructions when the encoder becomes weaker and the decoder starts ignoring the latent vectors. Additionally, learning a continuous latent space can be computationally expensive and is not necessary for certain tasks like sequential decision making.

VQ-VAE has a similar structure to that of an autoencoder, with an encoder network, a discretized latent space, and a decoder network. Unlike VAE, which uses a continuous latent space, VQ-VAE uses a discrete latent space to ensure that the encoder does not encode weak information and to improve optimization efficiency when working with sparse data with discretized nature. **Figure 1** shows the architecture of a VQ-VAE used by the authors [2, 3]. The indices of embedding vectors corresponding to each input sample are referred as the latent code. These latent codes are used by decoder to reconstruct the input.

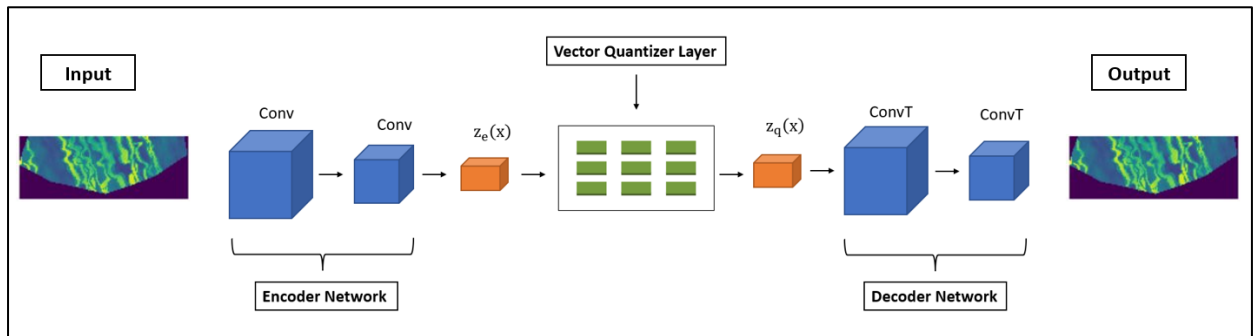

**Figure 1:** A schematic of the VQ-VAE architecture used in [2]. The encoder and decoder consist of convolutional networks. The VQ layer contains integer codes (referred as latent codes), with each one representing an embedding vector in the codebook.

The latent space is created through the use of a codebook that contains learnable embedding vectors. The encoder network learns to generate outputs that are as close as possible to a few selected embedding vectors in the codebook, which are referred to as the latent codes. This assignment is done through a nearest neighbor lookup process known as Vector Quantization (VQ). The learning of the encoder network is guided by a commitment loss and a reconstruction loss, while the decoder network learns to reconstruct the input data by sampling a sequence of certain embedding vectors from the codebook. The decoder network learns by minimizing the reconstruction loss.

The reconstruction loss in VQ-VAE is used to tune both the encoder and the decoder. This loss measures the difference between the input data and the reconstructed data. The codebook loss, on the other hand, is used to maintain the proximity between the encoder output and the embedding vectors in the codebook. This loss ensures that the embedding vectors in the codebook are continuously updated to better align with the encoder output. The commitment loss in VQ-VAE is used to refine the encoder, keeping it from fluctuating too frequently from one code vector to another. This loss ensures that the encoder outputs remain committed to their assigned embedding vectors in the codebook.

### 1.1.4 Hierarchical Vector-Quantized Variational Autoencoder (VQ-VAE-2)

VQ-VAE-2 is organized in a hierarchical manner, with multiple layers of decoders, encoders, and vector quantization (VQ) layers [7]. For simplicity, it usually takes a two-layer architecture. The hierarchical structure of VQ-VAE-2 ensures better reconstruction compared to VQ-VAE. The hierarchical multi-scale latent representations increase the resolution of reconstruction. In a two-layer form VQ-VAE-2, the top latent code represents global information, while the bottom latent code represents the local details. The top code captures the global structure of spatial variations in the geomodel layer. The bottom code contains the local details and finer structural variations in the geomodel layer. Each component in the code has an integer value representing an index of an embedding vector in the codebook.

## 1.2 Geomodel Generation Techniques

Geostatistical modeling is a technique for generating geomodels using spatially correlated data in earth science. Geostatistical modeling methods, in general, can be categorized into Kriging/Kriging-based simulation and Multi-point statistics (MPS) simulation. An ensemble of geomodels are generated using these approaches by integrating pixel-level data or pattern-level objects. After the generation, the geomodels are tuned to match the observed field production data, which is known as history matching or data assimilation. Attempts have been made to calibrate geomodels based on flow responses by combining deep generative neural networks and history matching techniques [8,9,10]. However, this is computationally prohibitive in that forward simulation for thousands of realizations is needed. In this paper, we employ a pure data-driven method to assist realistic geomodel generation.

### 1.2.1 Kriging and Kriging-based Simulations

Kriging, developed by Matheron [11], is one of the most fundamental and widespread geostatistical techniques used in geoscience community. Ordinary Kriging requires a prior knowledge of variability and correlation of spatial data, referred as variogram, to produce the geomodel. It is a deterministic method and thus only generate one geomodel from the geo-data. Moreover, this method turns out to produce smoothing models and it cannot represent heterogeneity. To overcome these limitations and effectively quantify uncertainties, Kriging-based stochastic simulation is developed to produce an ensemble of equiprobable geomodels by estimating an unknown node each iteration following a random path through grid nodes [12]. These modeling approaches, however, fail to reproduce complex patterns such as flow channels because they rely on two-point statistics. Another downside of these approaches is the slow computational speed, especially when millions of grid cells are to be simulated.

### 1.2.2 Multi-point statistics (MPS)

Geomodel generation using Kriging method simulates one grid/cell at a time using two-point statistics; therefore, it cannot produce complex and realistic geological patterns [13]. Multiple-point statistics (MPS) method is proposed to help deal with complex geological features. It takes in training image (TI) or a groups of TIs as main input. TI is a conceptual representation of the complex geological systems such as facies, depositional structures and so on. MPS method can be categorized into two groups: pixel-based and pattern-based [13]. Geomodel generation using MPS method also suffer from high computational cost because of their sequential simulation nature.

### 1.2.3 Neural Network based Geomodel Generation

Attempts have been made to generate geomodels based on production data by combining deep generative neural networks and history matching techniques. Authors in [9] employs convolutional variational autoencoder (CVAE) and ensemble smoother to parameterize the facies geomodels. It outperforms traditional data assimilation techniques in terms of facies reconstruction. However, CVAE models are prone to reproduce unrealistic samples because it suffers posterior collapse [14]. Similar to CVAE, a generative adversarial network (GAN) was added with ensemble smoother to generate facies geomodels and showed decent performances [10]. A GAN based model, on the other hand, trains a generator and a discriminator simultaneously. Training the GAN is challenging because it requires finding a point of Nash equilibrium for two competing networks. Moreover, appropriate design of the network architecture is needed to avert mode collapse and instability.

### 1.2.4 Research Objectives

In this paper, we train VQ-VAE-2 followed by training of PixelSNAIL. Based on the principle of neural discrete representation learning, the VQ-VAE-2 learns to massively compress the training geomodels to extract the low-dimensional, discrete latent representation corresponding to each geomodel. Following that, PixelSNAIL uses deep autoregressive network to learn the prior distribution of the latent codes representing the training geomodels. The trained PixelSNAIL can sample from the latent-code distribution to generate new geomodels that preserve geological realism and consistency. To the best of knowledge, this is the first geomodel generation practice using deep autoregressive model and deep learning techniques in geoscience community. The benefits of using PixelSNAIL are twofold: Firstly, it gives explicit estimation of the probability density function of the priors. Secondly, training and sampling directly over the latent space is way faster than iterating over the cells of the large geomodels.

## 2 Description of the Dataset

The dataset contains porosity, permeability and other quantities of a synthetic oil field – Brugge field. The dimensions of the field are roughly 15km×5km and the geomodel consists of approximately 60000 grid cells, with an average cell dimension of 125m×125m×7m. When modeling the geological facies, an object-based stochastic simulation was used for the facies in the upper and lower zone. To quantify uncertainties, there are 5000 realizations being generated. Table 1 contains the geological and statistical summary of the properties used to generate the geomodels.

**Table 1:** Model Stratigraphy and average properties of Brugge model [2].

| Formation Type | Average net-to-gross ratio (%) | Average Porosity (%) | Average Perm (mD) | Average Saturation (%) | Layers | Facies Type | Deposition environment |
|---|---|---|---|---|---|---|---|
| Upper Zone | 83 | 18 | 445 | 47 | 1-5 | Channel Complex (2 reservoir facies) | Fluvial (Low Aspect ratio) |

| Lower Zone | 86 | 19 | 550 | 46 | 6-9 | Channel Complex (2 reservoir facies) | Fluvial (High Aspect ratio) |

## 3 METHODOLOGY

The implementation entails two steps: 1) Compress training geomodel realizations into hierarchical discrete latent representations with VQ-VAE-2, and 2) Generate new geomodels using PixelSNAIL model and learned decoder network. More explanations are detailed in the following subsections.

### 3.1 Geomodel Compression with VQ-VAE-2

In contrast to other dimensionality reduction techniques, vector quantized variational autoencoder-2 (VQ-VAE-2), because of its multiscale hierarchical architecture and quantization, achieves the highest compression ratio while maintaining decent reconstruction performance for geomodels. This was well elaborated and tested in an earlier publication by the authors [2]. In **Figure 2**, a schematic plot of two-level VQ-VAE-2 architecture is presented. It is basically hierarchical VQ-VAE with additional top encoder, top VQ layer and top decoder.

In this work, we stick to a two-level hierarchical architecture where the top latent code contains global information about the geomodel, while bottom latent code stores fine, local details of spatial variations. There are 2000 geomodels in the dataset. Each geomodel contains 9 layers. Each layer of geomodel contains 144x48 cells. Some hyperparameters of the network are listed in Table 2 below.

**Table 2:** Hyperparameters of the VQ-VAE-2 network with a compression ratio 167

| Top codebook size, $K_t$ | 1024 | Bottom codebook size, $K_b$ | 1024 |
|---|---|---|---|
| Top latent code dimension | 2x4 | Bottom latent code dimension | 4x8 |
| Top codebook vector length, $D_t$ | 64 | Bottom codebook vector length, $D_b$ | 64 |

Different compression ratios are achieved by changing latent code dimensions, e.g. down sampling, changing strides of Conv2D layer, etc. For compression ratio of 167, as shown in **Figure 2**, the dimension of top/bottom latent map is (2,4) / (4,8), as shown in Table 2. In other words, the top latent code contains 8 components, while the bottom latent code contains 32 components. For compression ratio of 417 and 667, the dimension of top/bottom latent map is (2,4) / (2,4) and (1,2) / (2,4), respectively. We refer readers to two past research publications [2, 7] for more details of the network architecture. However, because previous models use mean squared error for the loss layer, reconstructed models turn out to be blurry in regions of flow channels. For this reason, we propose to use perceptual loss to better reconstruct flow channels of geomodels, which will be discussed in the next section.

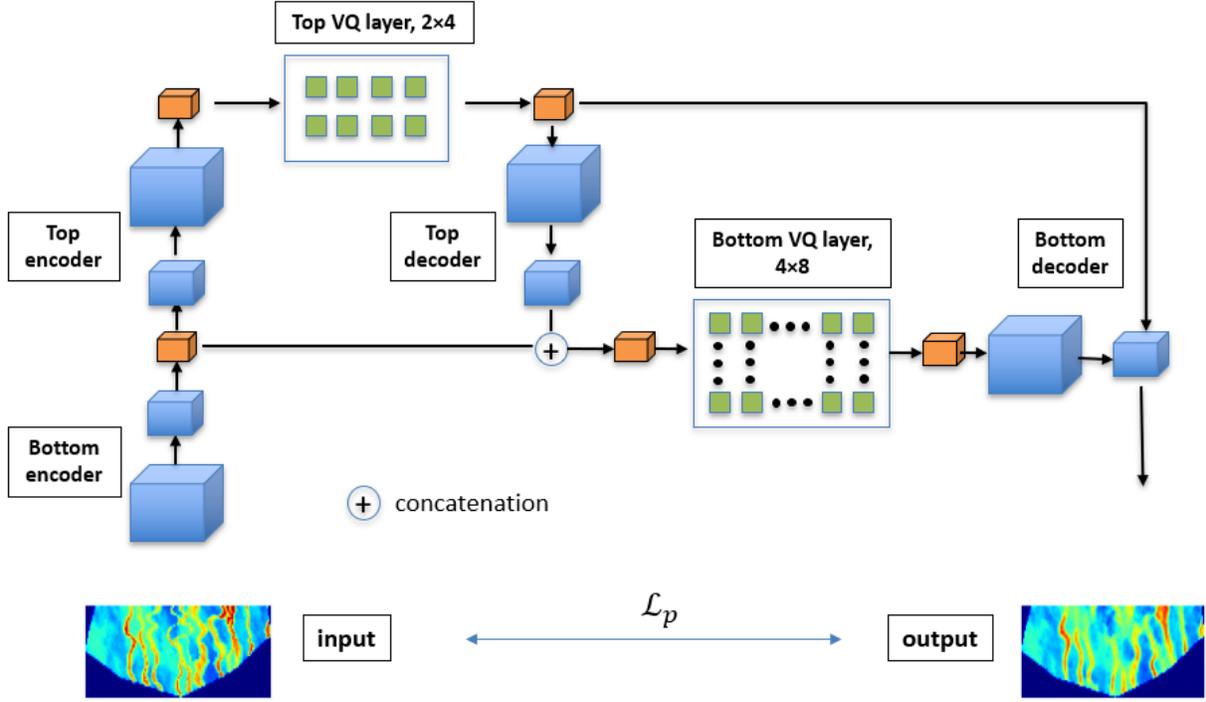

**Figure 2:** A schematic of the VQ-VAE-2 architecture for compression ratio 167. It is a hierarchical version of a traditional VQ-VAE, shown in Figure 1. The cubic box represents convolutional layers. $\mathcal{L}_p$ represents the model is trained using perceptual loss.

### 3.2   Perceptual Loss

For channelized geomodels, preserving flow channels is essential because flow channels provide tortuous yet high fluid transport pathways. Perceptual loss is a state-of-the-art loss metric that respects human's vision system with the consideration of image luminance, structure and contrast. Compared to per-pixel loss, perceptual loss is better in terms of producing perceptually desirable results.

Prevalent perceptual metrics include structural similarity index measure (SSIM) and multi-scale structural similarity index measure (MS-SSIM). SSIM is a similarity measure between two images, it can be factorized into three components: luminance $l$, structure $s$ and contrast $c$ measurements [15]. Luminance $l$ is used to compare the brightness of two images. Structure $s$ measures the correlation of the patterns in an image/geomodel, i.e. if a pixel value in image x is above its mean and so is the corresponding pixel in image y, then this is regarded as similar pattern of deviations, leading to similar structures. Contrast $c$ is a measure of how pixel values are spread in an image and is obtained using standard deviation of the image. The formula of SSIM between ground truth geomodel $x$ and reconstructed geomodel $y$ can be expressed as:

$$SSIM(x,y) = l(x,y)^\alpha \cdot c(x,y)^\beta \cdot s(x,y)^\gamma$$

where $\alpha, \beta, \gamma$ are weights that determines relative importance of three components. Generally, the three weights are set as 1.

MS-SSIM is a multi-scale version of SSIM. MS-SSIM is conducted at different resolutions through a multi-stage process of down-sampling the image to obtain various image resolutions [16]. MS-SSIM is formulated as:

$$\text{MS-SSIM}(x, y) = l_M(x,y)^{\alpha_M} \cdot \prod_{j=1}^{M} c_j(x,y)^{\beta_j} s_j(x,y)^{\gamma_j}$$

where $M$ represents the highest scale level.

However, SSIM and MS-SSIM alone may not produce good results. We use a mixed loss [17] that combines MS-SSIM and $l_1$ to achieve decent luminance and color while maintaining high contrast in high frequency regions. To better preserve structure and contrast of flow channels, we opt to select the mixed perceptual loss. The mixed perceptual loss is expressed as follows:

$$L^{mix} = \alpha \cdot L^{MS-SSIM} + (1 - \alpha) \cdot G_{\sigma_G^M} \cdot L^{l_1}$$

where $L^{MS-SSIM}$ is the MS-SSIM loss. Minimizing the total loss requires maximizing the MS-SSIM metric; so, $L^{MS-SSIM} = 1 - \text{MS-SSIM}$. $G_{\sigma_G^M}$ denotes the standard deviation of Gaussian filter, $L^{l_1}$ represents $l_1$ loss function, and $\alpha$ is a coefficient between 0 and 1, which is set at 0.80 in our study. $l_1$ loss function is simply the mean absolute error loss which is defined as follows,

$$L^{l_1}(P) = \frac{1}{N} \sum_{p \in P} |x(p) - y(p)|$$

where $p$ is pixel index, $P$ is the whole patch of ground truth geomodel $x$ and reconstructed geomodel $y$.

### 3.3 Compression of Multi-attribute Geomodels

In our study, a single-attribute geomodel represents the 3D spatial distribution of one property. For compression, a single-attribute 3D geomodel is split into 2D horizontal layers. Each 2D layer represents the spatial distribution of one property. Each 2D layer is repeated 3 times to create the input to the VQ-VAE-2 for compression and reconstruction. Each repetition acts as one of the 3 channels that is fed into the VQ-VAE-2. Each channel is the same 2D layer representing the same spatial distribution of one property. For single-attribute demonstration in the paper, we focus on the spatial distribution of porosity.

In our study, a multi-attribute geomodel represents the 3D spatial distribution of three properties, namely porosity, permeability and water saturation. For compression, a multi-attribute geomodel is split into 2D layers, where each 2D layer represents the spatial distribution of one property. The VQ-VAE-2 model compresses and reconstructs three 2D layers, representing 3 properties, simultaneously. Each 2D layer is treated as one of the three channels that is processed by the VQ-VAE-2 model. Each of the three channel represents one of the three properties, namely porosity, permeability and water saturation. The single-attribute and multi-attribute geomodel compression use the same VQ-VAE-2 network architecture.

### 3.4 Geomodel Generation with PixelSNAIL Model followed by the VQ-VAE-2 Decoder Network

Unlike traditional geomodel generation approaches, e.g. Kriging, MPS. Our approach requires a few representative prior geomodel realizations for training. Prior population of training geomodels are generated using object-based stochastic simulation. Our approach implicitly honors hard data (e.g. rock property data at certain locations) because the information is embedded in the latent space representations when training the network. In our geomodel generation method, generating new geomodels is as simple as sampling latent codes from known distribution and decoding the latent codes using pretrained decoder network. To the best of knowledge, this is the first geomodel generation practice using autoregressive approaches.

#### 3.4.1 PixelSNAIL – A Deep Autoregressive Network

Deep generative modeling is a class of neural-network-based unsupervised learning that aims to generate new samples (e.g. image) by learning the distribution (probability density) of a training dataset. In order to generate, the training data distribution/density needs to be first learned. Generative models learn the probability density either explicitly or implicitly. Once the density is learnt, we can sample the implicit/explicit density to generate new synthetic data, similar to the training data.

Generative models are categorized based on how they model the probability density function of the data. Autoregressive methods provide a tractable explicit density of the data. Variational autoencoders provide an approximate non-tractable explicit density, while generative adversarial networks provide a direct implicit density. Autoregressive methods include, but not limited to, fully visible belief nets, neural autoregressive density estimator (NADE), pixel recurrent neural network (PixelRNN), and pixel convolutional neural network (PixelCNN). In this work, we use an autoregressive method similar to PixelCNN [18], referred as PixelSNAIL.

PixelCNN contains convolutional layers to increase receptive field in order to make the training faster. Masks are also used to circumvent violation of causality law. In PixelCNN, there are two types of masks, including mask A where the central pixel is zeroed out, and mask B where the central pixel is allowed to connect to the pixel in the next layer [18, 19]. PixelCNN suffers from blind spot problems in the receptive field, making the generation performance undesirable. Several variants of PixelCNN, such as Gated PixelCNN [20], PixelCNN++ [21] and PixelSNAIL [22] were developed to circumvent the blind spot problem and improve the generative modeling performance.

PixelSNAIL is a state-of-the-art PixelCNN model with casual convolution and self-attention layers, which enables the model to learn long-range dependencies of components in a sequence, e.g. pixels in image [22]. In our study, PixelSNAIL will learn the dependencies between the latent-code components. The architecture of PixelSNAIL network is presented in the Appendix B. It has been implemented in areas like medical image generation, 3D facial image synthesis, etc. For example, PixelSNAIL was utilized to synthesize 3D face images and the author demonstrates that generated face images are closer to real faces [23]. PixelSNAIL was also used to increase the database of microscopic skin lesion and shows that it is a powerful technique in generating lesion images whilst retaining an easy-to-train property [24].

In our study, PixelSNAIL learns the prior distribution of the latent codes in a self-supervised fashion, with its model parameters tuned by maximizing likelihood of reconstructing the training latent codes as the output of the PixelSNAIL. The model's capability of generating new images comes from sampling latent codes from learned explicit conditional probability functions. The architecture of PixelSNAIL network for learning the prior distribution of top latent code is slightly different from that of the bottom latent code. Top code represents the global information and the bottom code represents the local and finer details. In the model for the top code, self-attention mechanism is used, while for the bottom code, large conditioning stacks are included because the bottom code should be sampled given the top code.

### 3.4.2   Workflow for Geomodel Generation

The geomodel generation in our study involves 4 steps. First, hierarchical latent codes (top and bottom codes) are computed for each training geomodel using the trained VQ-VAE-2 model. Second, the top latent codes of training geomodels computed using the trained VQ-VAE-2 encoder are passed into PixelSNAIL model for a self-supervised training of the top PixelSNAIL. The top PixelSNAIL learns the density of top latent codes. In the next stage, bottom PixelSNAIL is trained using the bottom latent codes of the training geomodels, such that the self-supervised training of the bottom PixelSNAIL is conditioned on the top latent codes. The bottom PixelSNAIL learns the density of bottom latent codes. In Stage 4, the two trained PixelSNAIL models, top and bottom PixelSNAILs, are used to sample new top and bottom latent codes. The newly sampled latent codes are then passed into trained VQ-VAE-2 decoder to generate new geomodels. We demonstrate stage 4 with two specific examples in Appendix A. The workflow of geomodel generation is shown in **Figure 3**.  A sketch plot of pre-trained VQ-VAE-2 decoder is presented in **Figure 4**.

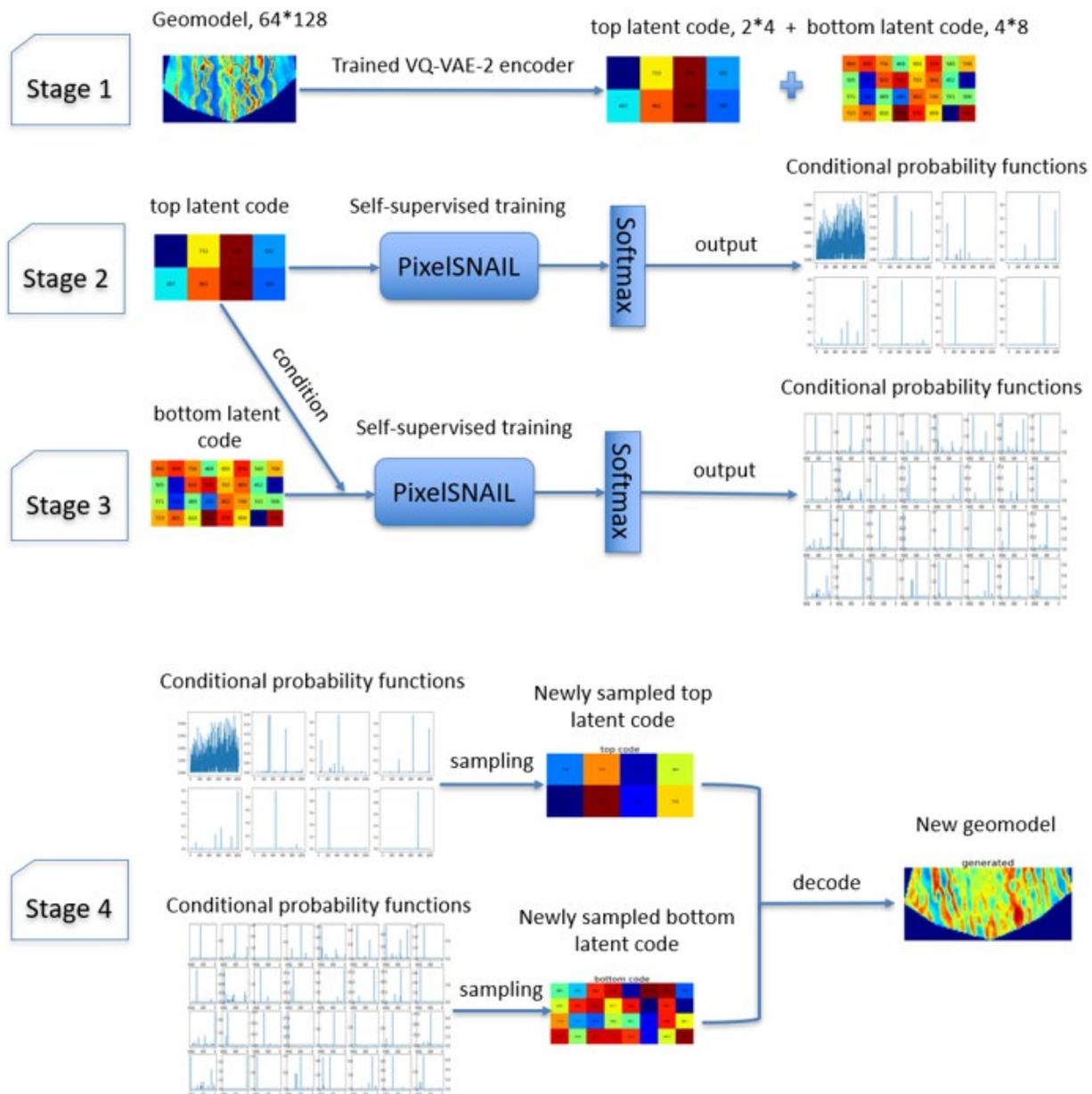

**Figure. 3.** Workflow for the unconditional geomodel generation. Stage 1: Hierarchical latent codes (top and bottom codes) are computed for each geomodel. Stage 2: The top latent codes of training geomodels are passed into PixelSNAIL model for a self-supervised training of the top PixelSNAIL. Stage 3: Bottom PixelSNAIL is trained using the bottom latent codes of the training geomodels conditioned on the top latent codes. Stage 4: The two trained PixelSNAIL models, top and bottom PixelSNAILs, are used to sample new top and bottom latent codes that are then passed into trained VQ-VAE-2 decoder to generate new geomodels.

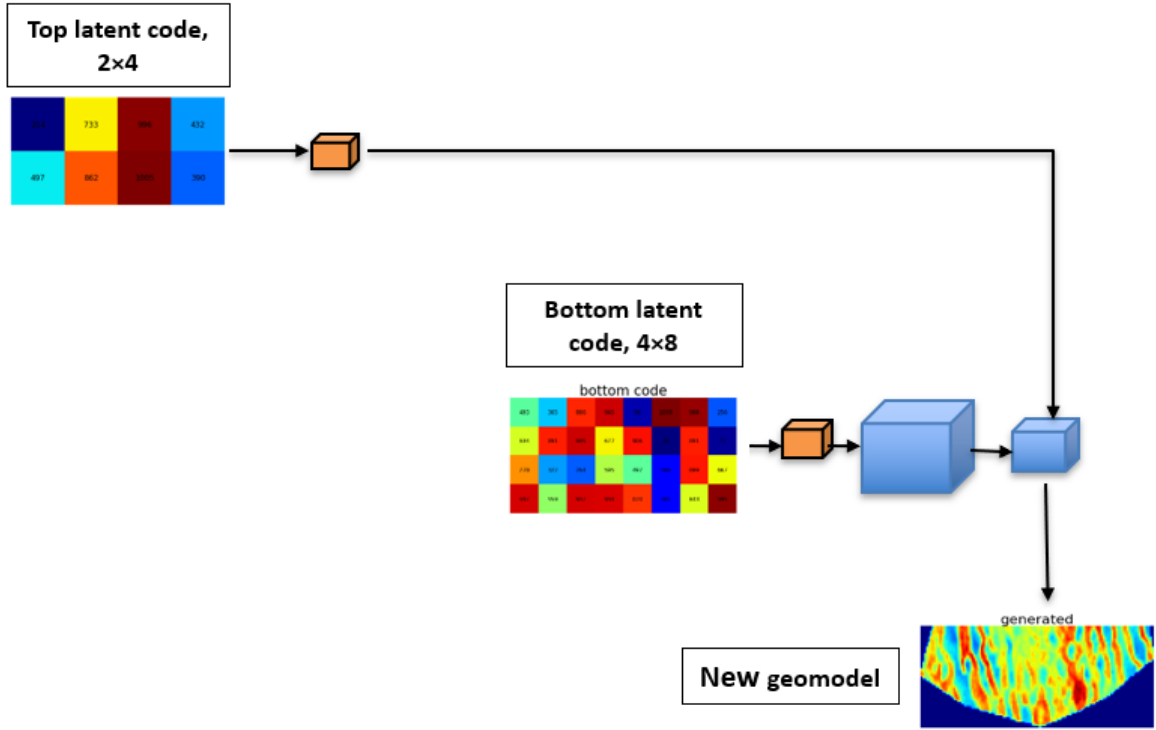

**Figure. 4.** VQ-VAE-2 decoder uses newly sampled top and bottom latent codes to generate new geomodel. This is the final step of the geomodel-generation workflow, shown in Figure 3. The two trained PixelSNAIL models, top and bottom PixelSNAILs, are used to sample new top and bottom latent codes that are then passed into trained VQ-VAE-2 decoder to generate new geomodels.

### 3.4.3 Sampling Codes using the trained PixelSNAIL

Due to its autoregressive nature, the sampling of the codes has to be sequential, that is to say, the 2D latent codes are generated component by component (i.e. code element by code element) from upper left to lower right. When sampling a component (i.e. code element), the sampling is conditioned on all previous elements/components in the code sequence. Moreover, the top code is sampled completely before sampling the bottom code because the bottom code is conditioned on the top code. The sampling procedure for the top latent code is as follows:

(i) Initialize the input to PixelSNAIL with all zeros. The dimension *m×n* of the initializing zero matrix equals to that of the top latent code, which is computed using the trained VQ-VAE-2.
(ii) The trained PixelSNAIL model processes the initializing zero matrix to generate a vector with $K_t$ elements ($K_t$ refers to the top codebook size which is defined in Table 2). Softmax layer then converts those numbers into a probability distribution vector with $K_t$ possible outcomes for the first component of the top latent code.
(iii) After sampling the first component from the probability distribution learnt in previous step, the input is reinitialized prior to computing the distribution of the second component.

(iv) As an iterative process, first compute the multinomial probability distribution of a component; then sample a value of the current component from the computed probability distribution.
(v) Re-initialize the input matrix by replacing with sampled value for the current component.
(vi) Go to the next component and repeat step iv-vi until all the components/elements of the top latent codes are sampled.

Sampling procedures of the bottom latent code are the same expect that the dimension of input zeros matrix is different and it should take in sampled top latent code as conditional input prior to computing the multinomial probability distribution.

### 3.4.4 Unconditional Geomodel Generation

Hundreds of geomodel realizations are generally created to reliably quantify the geological uncertainties of the subsurface. Unconditional generation of geomodels can be used as a data augmentation technique to produce new coherent and realistic geomodels based on the information present in the training dataset. In unconditional generation, the generation process is not dependent on any other information besides the entire training data itself. PixelSNAIL followed by VQ-VAE-2 decoder generates new geomodels, wherein the PixelSNAIL samples from the learned prior distribution of the latent code and the VQ-VAE-2 generates the geomodel using the sampled top and bottom latent codes.

The joint distribution of the latent code can be decomposed into product of conditional distributions of each element/component in the latent code. Each subsequent code element (or component) is sampled based on the previous sequence of code elements in the latent code. The mathematical formulation of the sampled latent code is a product of the sampled latent code elements, which is expressed as:

$$p(x) = \prod_{i=1}^{m \times n} p(x_i | x_1, \dots, x_{i-1})$$

where $m, n$ represents the height and width of the latent codes.

Unconditional generation can be combined with data assimilation approaches (Ensemble Kalman Filter and Ensemble Smoother) to help parameterize geomodels that preserve geological realism [9], where a convolutional variational autoencoder (CVAE) is integrated with ensemble smoother with multiple data assimilation (ES-MDA).

### 3.4.5 Conditional Geomodel Generation

In order to have control over the generation process, conditional generation was implemented in our study. Conditional generation in our paper is inspired by conditional PixelCNN application [20]. The mathematical formulation of the sampled latent code is a product of the sampled latent code elements conditioned on a specific latent code, which is expressed as:

$$p(x|h) = \prod_{i=1}^{m \times n} p(x_i | x_1, \dots, x_{i-1}, h)$$

where $h$ denotes the top and/or bottom latent code of the source geomodel, which is computed using the trained VQ-VAE-2 encoder. $h$ is passed through an unmasked convolution before adding to the activations. Specifically, in our study, $h$ is bottom latent code that stores local features of source geomodel, which is illustrated in Figure 5.

Conditional generation differs from unconditional generation in the following ways, which is further elaborated in Figure 5:

(i) The model input should take in source geomodel information as a condition when fitting and sampling of the top latent map
(ii) The model input should take in both top latent code and source geomodel information as conditions when training and sampling of the bottom latent code.

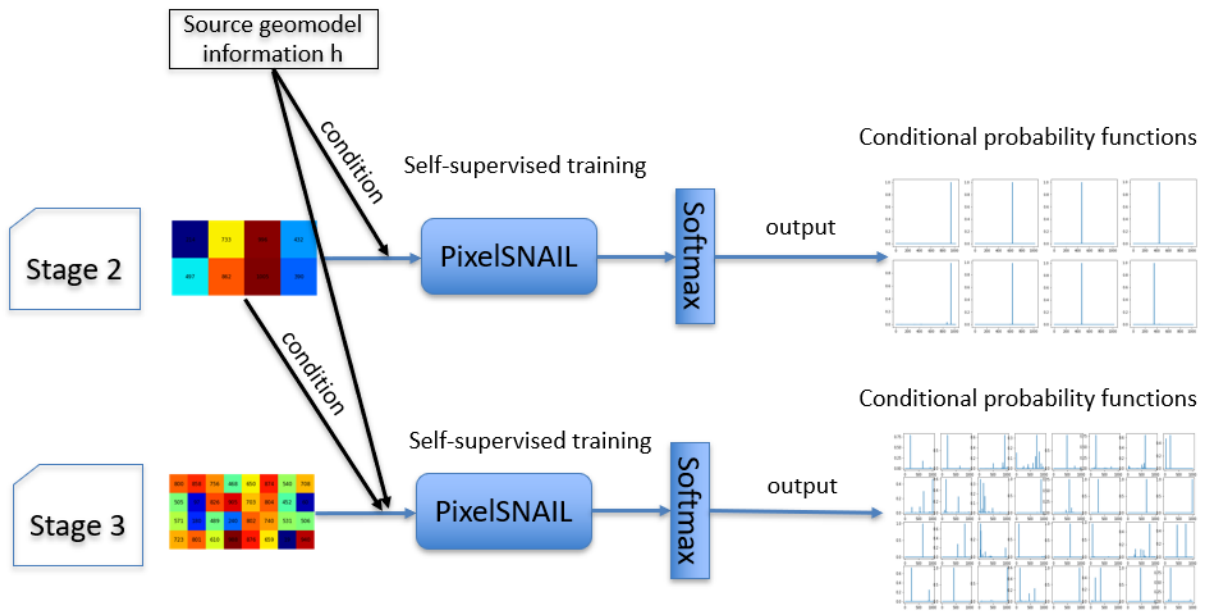

**Figure. 5.** Workflow for the conditional geomodel generation. Stages 1 and 4 are similar to those shown in Figure 3 and explained in Section 3.4.2. Both the top and bottom PixelSNAILs are conditioned on the source geomodel, h.

## 4 RESULTS AND DISCUSSIONS

### 4.1 Compression and Reconstruction with Mixed Perceptual loss compared to MSE loss

Table 3 compares the reconstructions from VQ-VAE-2 model with MSE loss against those with mixed MS-SSIM and $l_1$ loss at a fixed compression ratio 167. The top row of Table 3 shows original ground truth geomodels with increasing geological complexity from left to right. The middle row of Table 3 presents reconstructed geomodels using MSE loss, and the third row of the table displays reconstructed geomodels using mixed MS-SSIM and $l_1$ loss. As can be observed in Table 3, the VQ-VAE-2 model with MS-SSIM+$l_1$ can achieve better contrast in the flow channel regions; this is evident in the example presented in the last column of Table 3 (complex channels). The last row shows the color bar used to represent the porosity variation in the geomodels. Red-colored regions are the high-porosity channels with better transport properties, while blue-colored region represents the background reservoir with lower transport properties. All porosity illustrations in the paper use the same color scale.

**Table 3:** VQ-VAE-2 reconstructions of single-attribute geomodels with MSE loss versus those with MS-SSIM+L1 loss at a compression ratio 167

| | | | | |
|---|---|---|---|---|
| Original Geomodel | 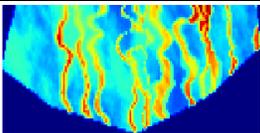 | 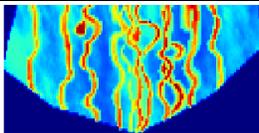 | 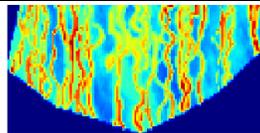 | 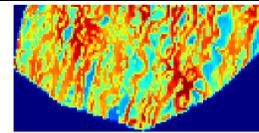 |
| VQ-VAE-2 with MSE loss | 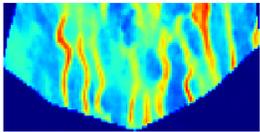 | 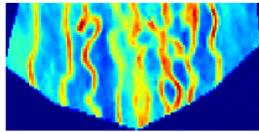 | 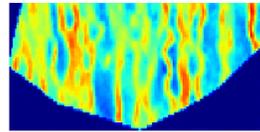 | 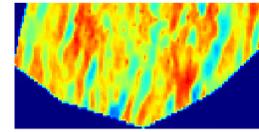 |
| VQ-VAE-2 with MS-SSIM + $l_1$ loss | 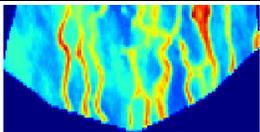 | 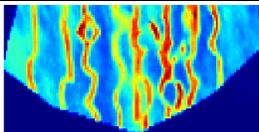 | 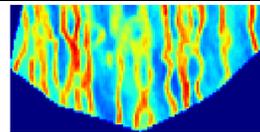 | 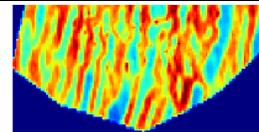 |
| Colorbar | 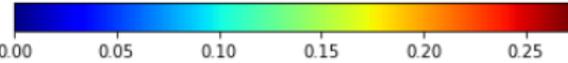 | | | |

## 4.2 Comparison of Compression and Reconstruction Performances between Single-attribute and Multi-attribute Geomodels

This section compares the reconstruction quality of VQ-VAE-2 at different compression ratios (CRs) for single-attribute geomodel compression (Table 4) against multi-attribute geomodel compression (Tables 5 and 6). In a single-attribute geomodel compression (Table 4), only one geomodel property, e.g. porosity, is compressed to a latent-code vector using the encoder and then the single property of the geomodel is reconstructed using the decoder. In a multi-attribute geomodel compression (Tables 5 and 6), multiple geomodel properties, such as porosity, permeability and saturation, are simultaneously compressed to one single latent code vector using one encoder, which is then reconstructed to generate all the properties of the geomodel using one decoder. These differences have also been explained in Section 3.3.

Tables 4, 5, and 6 present reconstruction performances at different compression ratios for 4 different testing examples, with increasing geological complexity from left to right. In Tables 4, 5 and 6, the first column shows the compression ratio, and the second column displays the mean and standard deviation (std) of the reconstruction in terms of SSIM, which quantifies the quality of reconstruction. Intuitively, a gradual increase of compression ratio from 104 to 617 leads to a decrease in average SSIM of reconstruction quality. Increase in SSIM to a value of 1 indicates perfect reconstruction.

**Table 4:** Porosity reconstruction at different compression ratios (CRs) for single-attribute geomodel compression using VQ-VAE-2. Four geomodels are evaluated in this Table. The geomodels are arranged in order of increasing geological complexity from left to right. The first column shows the compression ratio (CR). The second column displays the mean and standard deviation (std) of the reconstruction in terms of SSIM. Red-colored regions are the high-porosity channels with better transport properties, while blue-colored region represents the background reservoir with lower transport properties.

| CR | Mean and std of SSIM of reconstruction | Single-attribute Geomodel Compression and Reconstruction of Porosity | | | |
|---|---|---|---|---|---|
| **Original Geomodel** | 1.0000, 0.0000 | 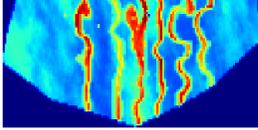 | 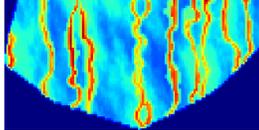 | 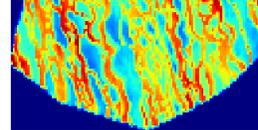 | 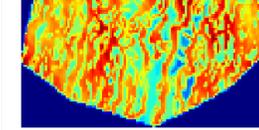 |
| **104** | 0.9466, 0.0116 | 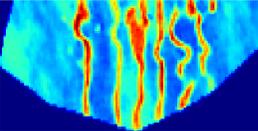 | 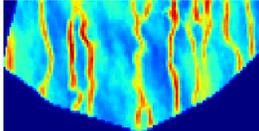 | 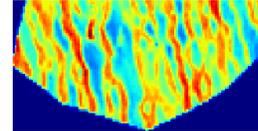 | 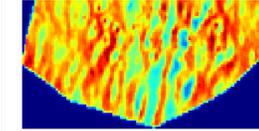 |
| **167** | 0.9260, 0.0145 | 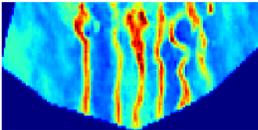 | 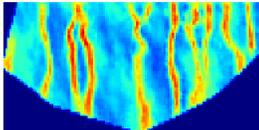 | 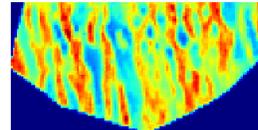 | 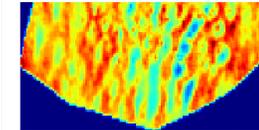 |
| **417** | 0.8999, 0.0192 | 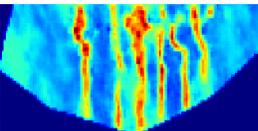 | 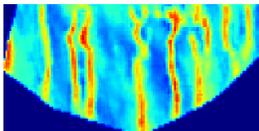 | 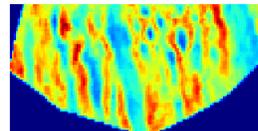 | 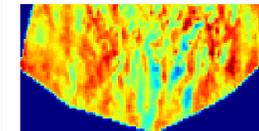 |
| **667** | 0.8931, 0.0205 | 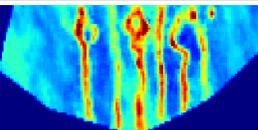 | 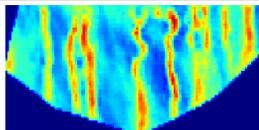 | 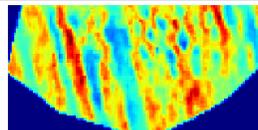 | 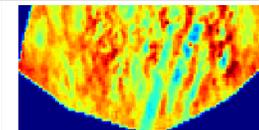 |
| Colorbar | | 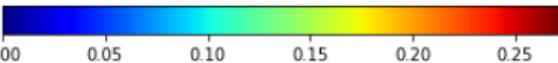 | | | |

Table 4 summarizes reconstruction performances of single-attribute geomodel compression for porosity. Tables 5 and 6 summarize reconstruction performances of permeability map and porosity map for the multi-attribute geomodel compression, respectively. At compression ratio 167 and 417, the multi-attribute geomodel achieves an average SSIM of 0.95 and 0.92, while single-attribute porosity geomodel has average SSIM of 0.93 and 0.90. Overall, at a specific compression ratio, the quality of reconstruction from multi-attribute geomodel compression is better than that from the single-attribute geomodel compression. The standard deviation of the reconstruction quality over the entire geomodel increases with increase in the compression ratio.

**Table 5:** Permeability reconstruction at different compression ratios (CRs) for multi-attribute geomodel compression using VQ-VAE-2. Four geomodels are evaluated in order of increasing geological complexity from left to right. The first column shows the compression ratio (CR). The second column displays the mean and standard deviation (std) of the reconstruction in terms of SSIM. Red-colored regions are the high-permeability channels with better transport properties, while blue-colored region represents the background reservoir with lower transport properties.

| CR | Mean and std of SSIM of reconstruction | Multi-attribute Geomodel Compression and Reconstruction of Permeability presented as log(perm+0.01) | | | |
|---|---|---|---|---|---|
| | | Testing geomodel 1 | Testing geomodel 16 | Testing geomodel 12 | Testing geomodel 21 |
| **Original Geomodel** | 1.0000, 0.0000 | 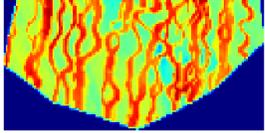 | 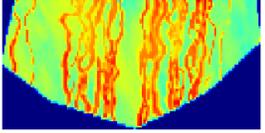 | 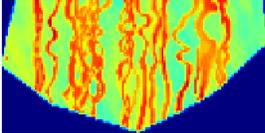 | 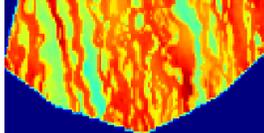 |
| **167** | 0.9258, 0.1704 | 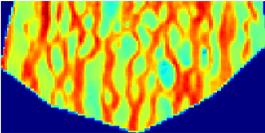 | 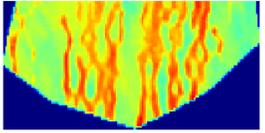 | 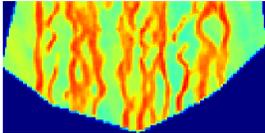 | 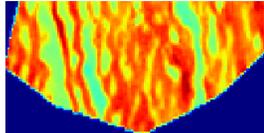 |
| **417** | 0.8611, 0.2281 | 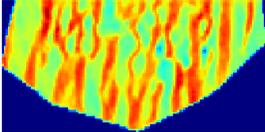 | 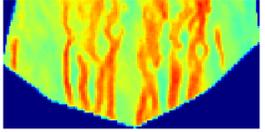 | 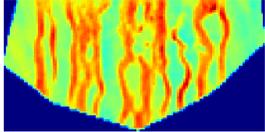 | 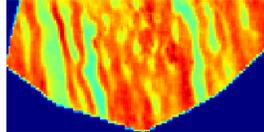 |
| **834** | 0.8162, 0.2205 | 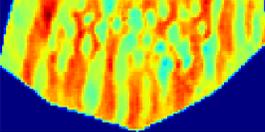 | 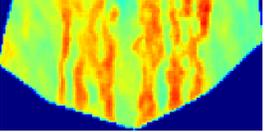 | 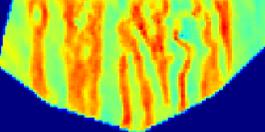 | 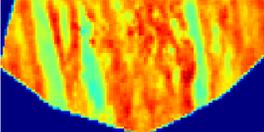 |
| Colorbar for log(perm+0.01) | | 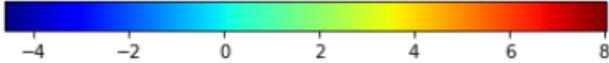 | | | |

**Table 6:** Porosity reconstruction at different compression ratios (CRs) for multi-attribute geomodel compression using VQ-VAE-2. Four geomodels are evaluated in order of increasing geological complexity

from left to right. The first column shows the compression ratio (CR). The second column displays the mean and standard deviation (std) of the reconstruction in terms of SSIM. Red-colored regions are the high-porosity channels with better transport properties, while blue-colored region represents the background reservoir with lower transport properties.

| CR | Mean and std of SSIM of reconstruction | Multi-attribute Geomodel Compression and Reconstruction of Permeability | | | |
|---|---|---|---|---|---|
| | | Testing geomodel 1 | Testing geomodel 16 | Testing geomodel 12 | Testing geomodel 21 |
| Original Geomodel | 1.0000, 0.0000 | 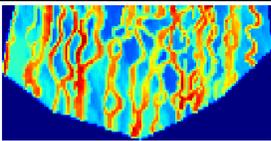 | 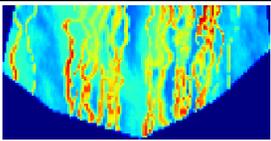 | 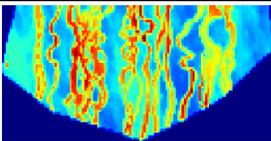 | 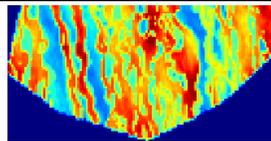 |
| 167 | 0.9536, 0.0104 | 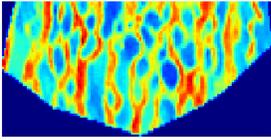 | 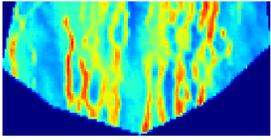 | 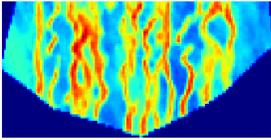 | 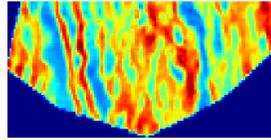 |
| 417 | 0.9211, 0.0154 | 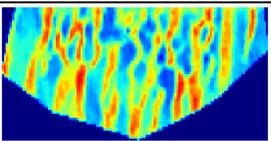 | 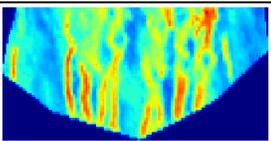 | 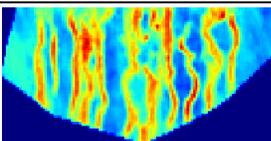 | 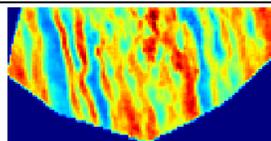 |
| 834 | 0.8759, 0.0263 | 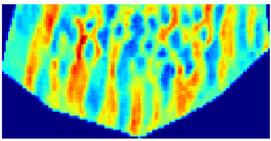 | 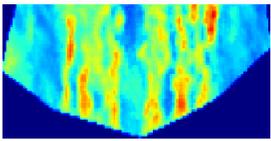 | 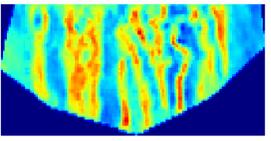 | 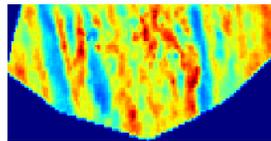 |
| Colorbar for porosity | | 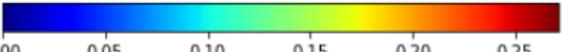 | | | |

### 4.3 Unconditional Geomodel Generation

**Figure 6** presents 20 newly generated geomodels with the unconditional PixelSNAIL followed by the VQ-VAE-2 decoder. These new geomodels are different than any of those from the training dataset while preserving the geological realism learnt from the training dataset. Moreover, the number, complexity and directionality of channels in the geomodels vary considerably from the geomodels available in the training dataset. The variation in generated geomodels increases model diversity, which is indispensable for geomodel uncertainty quantification and history matching. Generating new geomodels is way more efficient compared to traditional geomodelling approaches because generation is as simple as sampling latent codes from learned distribution and decoding sampled latent codes using pretrained VQ-VAE-2 decoder. The computation time for generating 300 geomodels is 56 seconds, which is run on a processor with Intel(R) Xeon(R) W-2333 CPU @ 3.60 GHz.

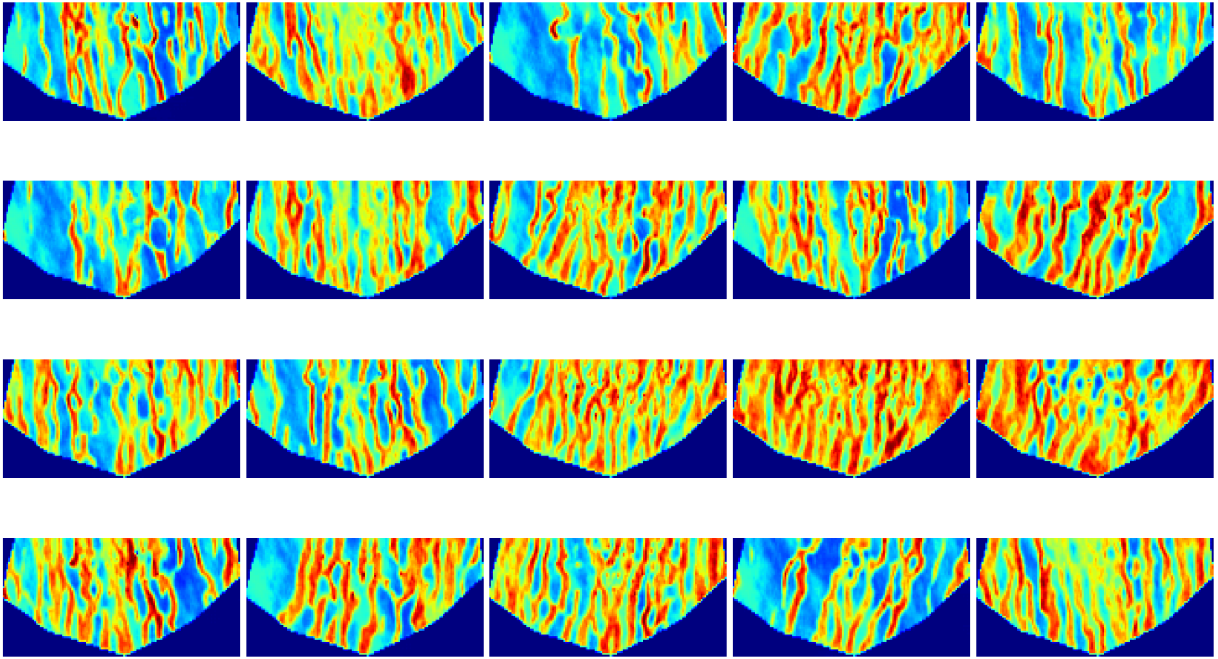

**Figure 6.** Unconditional geomodel (porosity) generation using the unconditional PixelSNAIL followed by the VQ-VAE-2 decoder. Red-colored regions represent high and blue-colored region represents low porosity. All porosity illustrations in the paper use the same color scale as shown in Tables 3, 4 and 6.

### 4.4 Conditional Geomodel Generation

**Table 7** presents conditional generation with the conditional PixelSNAIL followed by the VQ-VAE-2 decoder for 4 different source geomodels, presented in the first column. 300 geomodels were generated and ranked for each source geomodel. The ranking of a newly generated geomodel is based on the spatial and statistical comparisons between the source geomodel and the generated geomodel. The generated geomodels with most similarity to source geomodel will score lower and receive a better rank. For each generated geomodel, the scoring metric in our case is expressed as follows:

$$score = \left\{ \frac{1}{N_p} \sum_p^{N_p} \frac{1}{N_{reg}} \sum_i^{N_{reg}} |X_{i,p} - X_{i,p}^*| \right\} \times KL(X, X^*)$$

where $N_p$ is the number of statistical properties to be used in our metric for the region-specific comparison. We use mean, standard deviation, 25th percentile, 50th percentile and 75th percentile to compare the similarity of a selected sub-region. $N_{reg}$ is the number of subregions we partition the geomodel into for the region-specific comparison. $|X_{i,p} - X_{i,p}^*|$ is the absolute different between the generated geomodel and the source geomodel for a subregion *i* for a specific statistical property *p*. $KL(X, X^*)$ is the KL divergence between generated geomodel and source geomodel. Kullback-Leibler divergence was incorporated into the scoring metric to account for the differences in the property distribution. To be noted, the KL divergence is a non-spatial metric.

For each source geomodel, eight geomodels out of the 300 generated geomodels are displayed in **Table 7**. In the conditional generation of geomodel (constrained by the source geomodel), the generated geomodels exhibit geological consistency with the source geomodel with a limited spatial variation. As can be observed from Table 7, generated geomodels are respecting the spatial structure of the source geomodel. When conditioning the generation on source geomodels of lower geological complexity (e.g. the first one and last one in Table 7), the generated geomodels are similar to the source geomodel with little variation in channel numbers. When conditioning the generation on source geomodels of higher geological complexity (e.g. middle 2 examples in Table 7), the generated geomodels are also similar to provided source geomodel with little variation in channel directionality. This approach, although at the expense of losing model diversity, is a convenient way to control the geomodel generation. In doing so, we can use these generated geomodels for computing a large range of production scenarios due to variations in channel length and interconnectivity in the generated geomodel as compared to the source geomodel. Notably, the regions of high and low transport properties are preserved overall in these generations.

**Table 7:** Conditional geomodel generation using the conditional PixelSNAIL followed by the VQ-VAE-2 decoder. The source geomodels are provided in the first column. 8 newly generated geomodels per source geomodels are presented next to the source geomodel. The ranks indicate the similarity of the generated geomodels to the source geomodel. The ranks were assigned based on the scoring metric used to compare the generated and source geomodels. Red-colored regions represent high porosity and blue-colored region represents low porosity. All porosity illustrations in the paper use the same color scale as shown in Tables 3, 4 and 6.

| Source Geomodel | Newly Generated Geomodels Conditioned on the Source Geomodel | | | |
|---|---|---|---|---|
| (1) 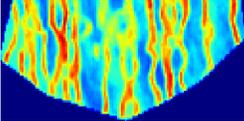 | 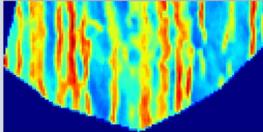 Rank: 1, score: 0.01057 | 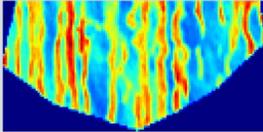 Rank: 9, score: 0.01417 | 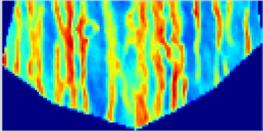 Rank: 19, score: 0.01417 | 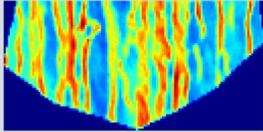 Rank: 29, score: 0.01603 |
| | 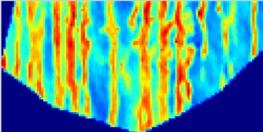 Rank: 39, score: 0.01681 | 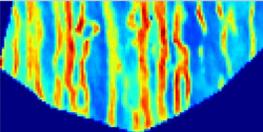 Rank: 49, score: 0.01749 | 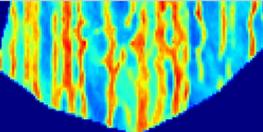 Rank: 59, score: 0.01813 | 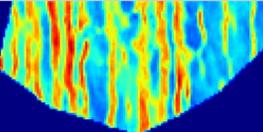 Rank: 69, score: 0.01847 |
| (2) 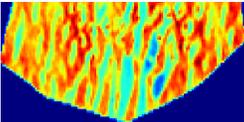 | 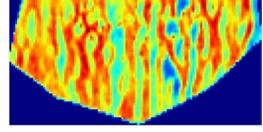 Rank: 1, Score: 0.01041 | 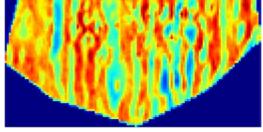 Rank: 9, Score: 0.01307 | 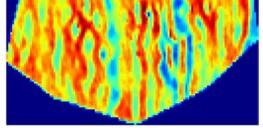 Rank: 19, Score: 0.01451 | 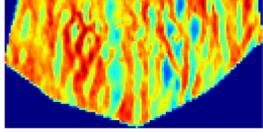 Rank: 29, Score: 0.01527 |
| | 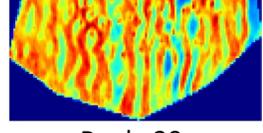 Rank: 39, Score: 0.01569 | 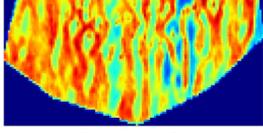 Rank: 49, Score: 0.01649 | 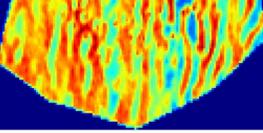 Rank: 59, Score: 0.01711 | 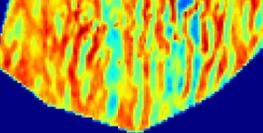 Rank: 69, Score: 0.01758 |

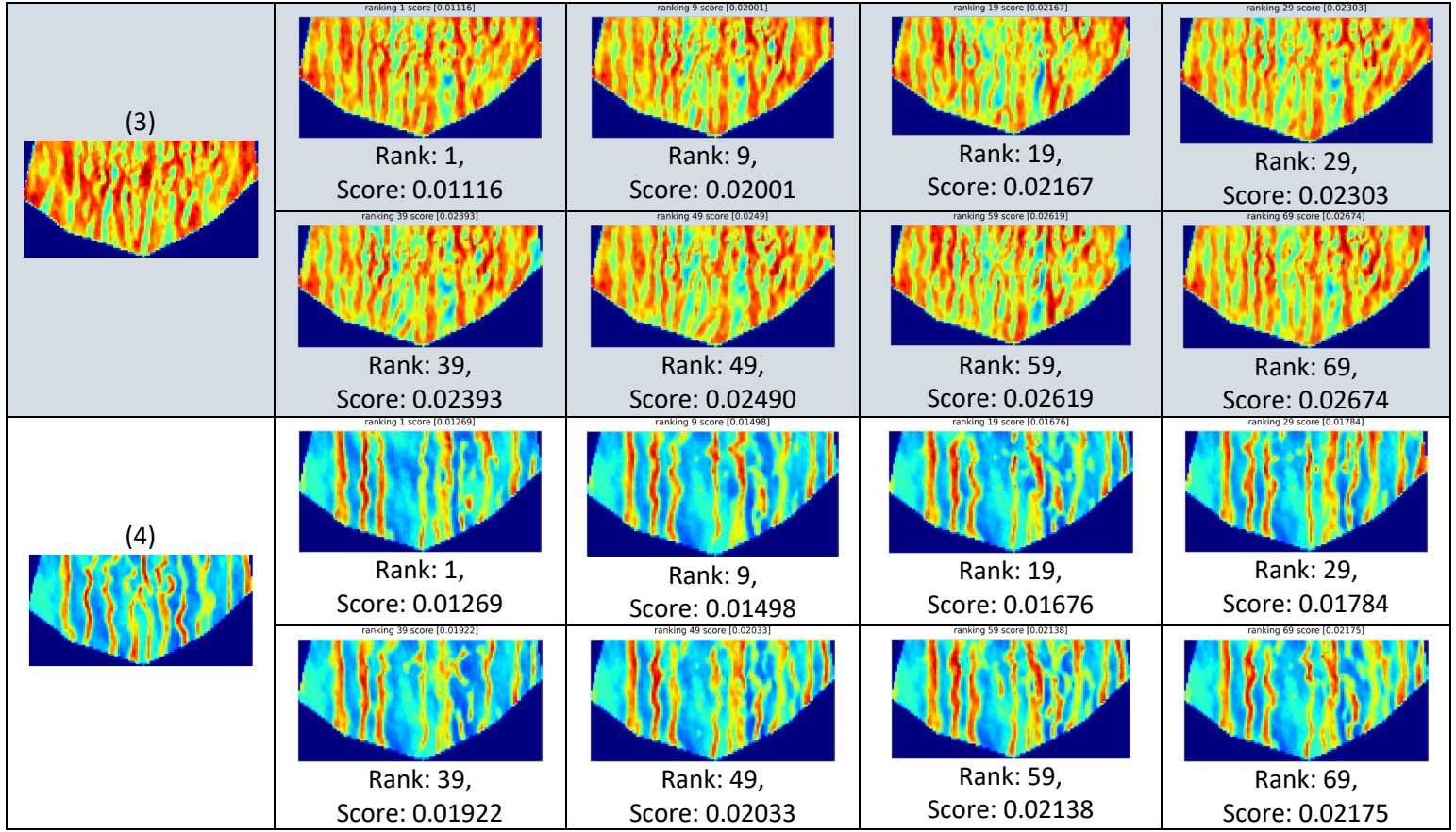

## 5 CONCLUSIONS

Massive compression of a large geomodel using vector-quantized variational autoencoder (VQ-VAE) produces extremely low-dimensional geological representation, referred as an integer-valued latent code. Such low-dimensional representations significantly reduce the storage requirements and computational cost of history matching, production forecast, and synthesis of new geomodels that ultimately benefits the subsurface characterization for geothermal energy, mining, carbon storage, ground water, and hydrocarbon recovery.

In this study, hierarchical VQ-VAE (referred as VQ-VAE-2) is implemented for extracting extremely low-dimensional representation of large geological model. VQ-VAE-2's hierarchical vector-quantization layers ensure higher resolution of reconstruction compared to the traditional VQ-VAE. By replacing the L2 loss with the perceptual loss, the reconstruction of tortuous, highly-conductive, channelized regions improves, especially in terms of contrast and boundary delineation. Higher compression ratio is achieved by stacking the permeability, porosity, and saturation of a multi-attribute geomodels as the three RGB channels, respectively, that are fed to the VQ-VAE-2 encoder network.

PixelSNAIL, a deep autoregressive network, learns to sample top and bottom latent codes from a tractable explicit density. PixelSNAIL followed by VQ-VAE-2 decoder generates new geomodels, wherein the PixelSNAIL samples from the learned prior distribution of the latent code and the VQ-VAE-2 generates the geomodel using the sampled top and bottom latent codes. The computation time for the unconditional generation of 300 geomodels is 56 seconds, which is run on a processor with Intel(R) Xeon(R) W-2333 CPU @ 3.60 GHz. This approach can be used to generate unique geomodels that varies considerably in terms of the number, complexity and directionality of channels. To generate geomodels with specified spatial patterns (in our case is the pattern from source geomodel), we employ conditional generation, where the model takes in conditional information from the source image to generate new geomodels constrained by the source geomodel. By designing score metric that measure the similarity between generated models and source models, we can select most similar models or least similar models. The selection is based on the tradeoff between model diversity and model similarity.

In summary, based on the principle of neural discrete representation learning, the VQ-VAE-2 learns to massively compress the training geomodels to extract the low-dimensional, discrete latent representation corresponding to each geomodel. Following that, PixelSNAIL uses deep autoregressive network to learn the prior distribution of the latent codes representing the training geomodels. The trained PixelSNAIL can sample from the latent-code distribution that can be fed to the VQ-VAE-2 decoder to generate new geomodels that preserve geological realism and consistency. To the best of knowledge, this is the first geomodel generation practice using deep autoregressive model and deep learning techniques in geoscience community.

**ACKNOWLEDGEMENTS**

This work was supported by ConocoPhillips through the Data-Driven Intelligent Engineering and Characterization (DICE) research group. Special thanks to Gaoming Li for providing valuable inputs to the project. We would also thank Yusuf Falola for his previous geomodel compression work. The work would not have been possible without Gustavo Gomez and John Hand, who initiated and have been supporting the project.

Seventh Asilomar Conference on Signals, Systems & Computers, 2003.

[17]   Zhao, H., et al. (2017). "Loss Functions for Image Restoration With Neural Networks." IEEE Transactions on Computational Imaging 3(1): 47-57.

[18]   Van Den Oord, A., et al. (2016). Pixel recurrent neural networks. International conference on machine learning, PMLR.

[19]   Pinaya, W. (2020, March). Autoregressive Models-PixelCNN. . https://towardsdatascience.com/autoregressive-models-pixelcnn-e30734ede0c1

[20]   Van den Oord, A., et al. (2016). "Conditional image generation with pixelcnn decoders." Advances in neural information processing systems 29.

[21]   Salimans, T., Karpathy, A., Chen, X., & Kingma, D. P. (2017). Pixelcnn++: Improving the pixelcnn with discretized logistic mixture likelihood and other modifications. arXiv preprint arXiv:1701.05517.

[22]   Chen, X., et al. (2018). Pixelsnail: An improved autoregressive generative model. International Conference on Machine Learning, PMLR.

[23]   Gallucci, A., et al. (2022). Generating High-Resolution 3D Faces Using VQ-VAE-2 with PixelSNAIL Networks. Image Analysis and Processing. ICIAP 2022 Workshops, Cham, Springer International Publishing.

[24]   Alessio Gallucci, Nicola Pezzotti, Dmitry Znamenskiy, Milan Petkovic, "A latent space exploration for microscopic skin lesion augmentations with VQ-VAE-2 and PixelSNAIL," Proc. SPIE 11596, Medical Imaging 2021: Image Processing, 115962X (15 February 2021); doi: 10.1117/12.2580664

[25]   Jubb, T. (2020, May). Autoregressive Generative Models in Depth: Part 4. AIs Blog. https://thomasjubb.blog/autoregressive-generative-models-in-depth-part-4/

[26]   Vaswani, A., Shazeer, N., Parmar, N., Uszkoreit, J., Jones, L., Gomez, A. N., ... & Polosukhin, I. (2017). Attention is all you need. Advances in neural information processing systems, 30.
25

# APPENDIX A: CONDITIONAL PROBABILITIES OF THE LATENT CODE ELEMENTS

Conditional probabilities of each element/component in the top and bottom latent codes computed using PixelSNAIL for unconditional generation. The conditional probability functions (CPF) are obtained sequentially. CPF of an element/pixel is estimated using trained PixelSNAIL network once all previous elements/pixels of the latent code have been sampled. Note that the first pixel of top latent map is unconditioned, while the first pixel of bottom latent map is conditioned on the whole top latent map.

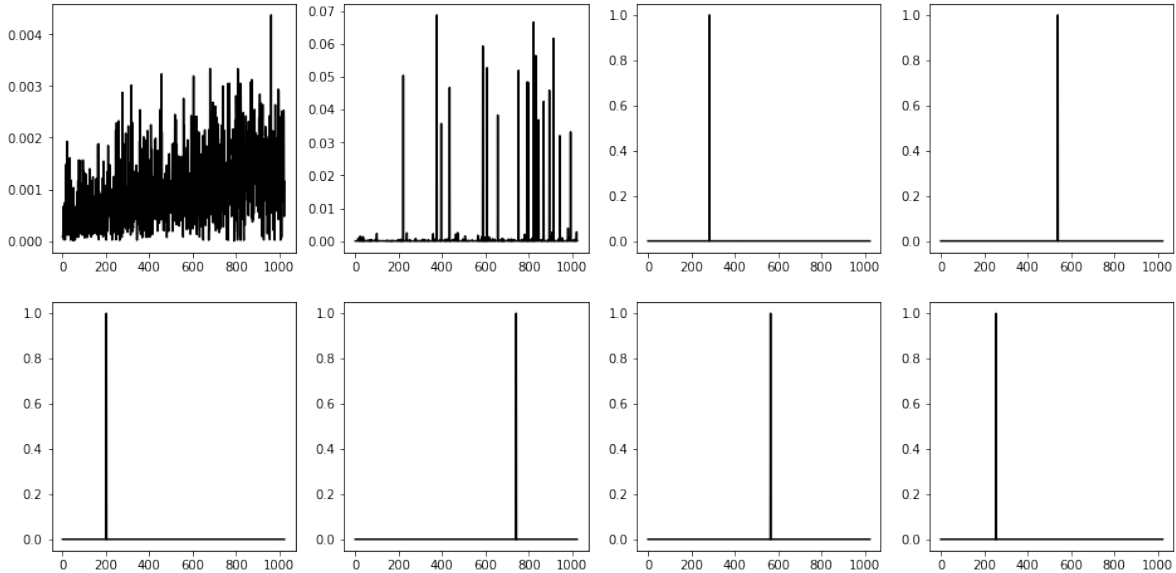

**Figure A1**. Conditional probability function of each of the 8 pixels of the top latent code that is used to sample the top latent code.

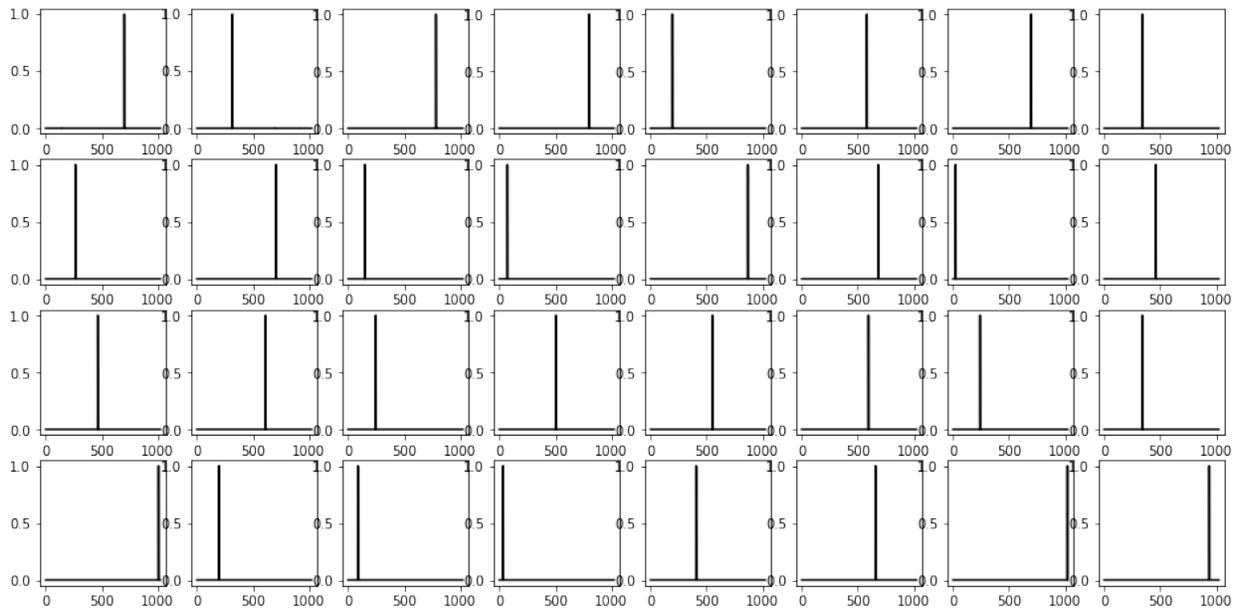

**Figure A2**. Conditional probability function of each of the 32 pixels in the bottom latent code that is used to sample the bottom latent code.

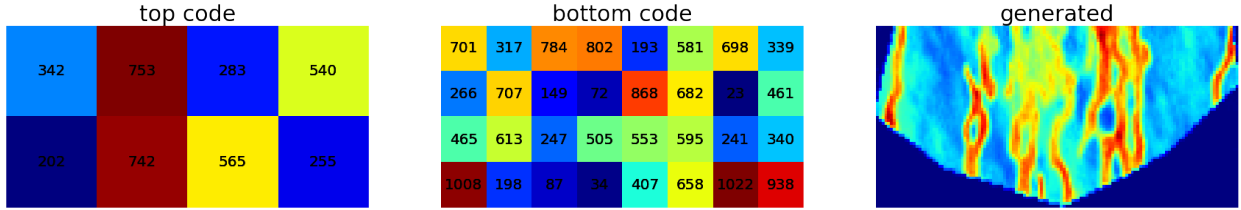

**Figure A3**. Newly sampled top and bottom latent codes obtained using the conditional probability functions shown in Figures A1 and A2. The numeric annotation denotes the latent code number, ranging from 0 to 1023. VQ-VAE-2 decoder uses the sampled top and bottom latent codes to generate a new geomodel, shown on the right.

Figures A1 and A2 show the conditional probability distribution from which the first set of top and bottom codes are sampled. Each subsequent element is conditioned on the previous sequence of elements in the latent code. The sampled top and bottom codes are shown in Figure A3. The newly generated geomodel using the first set of sampled top and bottom codes are shown on the right in Figure A3. Figures A4 and A5 show the conditional probability distribution from which the second set of top and bottom codes are sampled. The sampled top and bottom codes are shown in Figure A6. The newly generated geomodel using the second set of sampled top and bottom codes are shown on the right in Figure A6. Note that sampled latent codes in Figure A6 comes from the same probability distribution as in Figure A3, but they produce geomodels with totally different number of channels and different geomodel complexity.

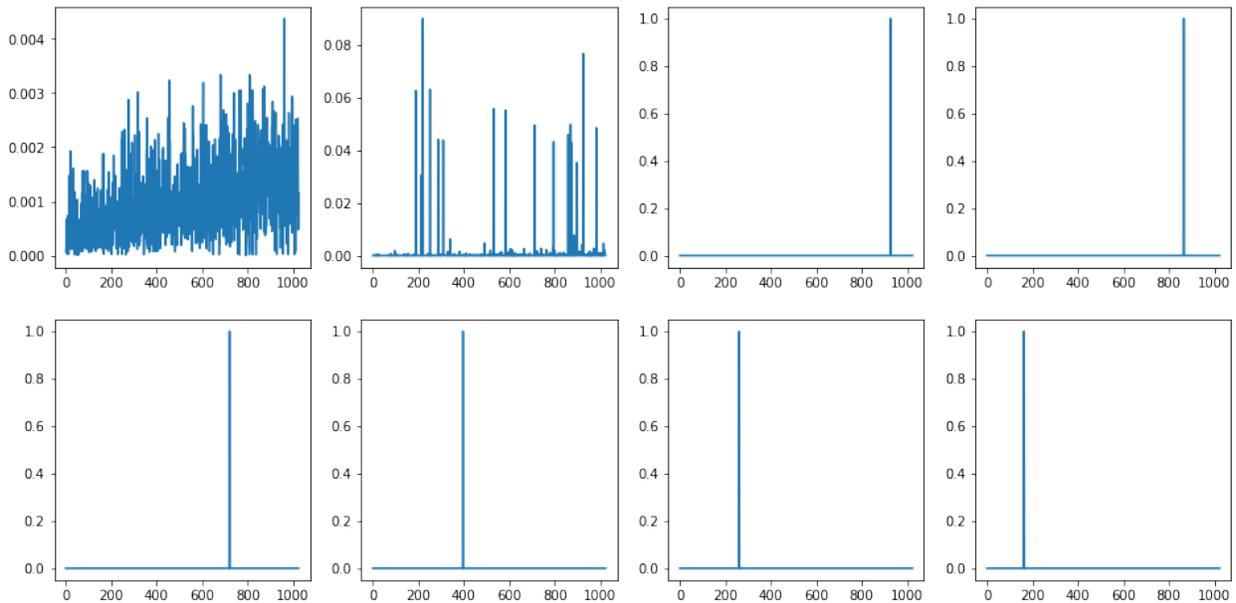

**Figure A4**. Conditional probability function of each of the 8 pixels of the top latent code that is used to sample the top latent code.

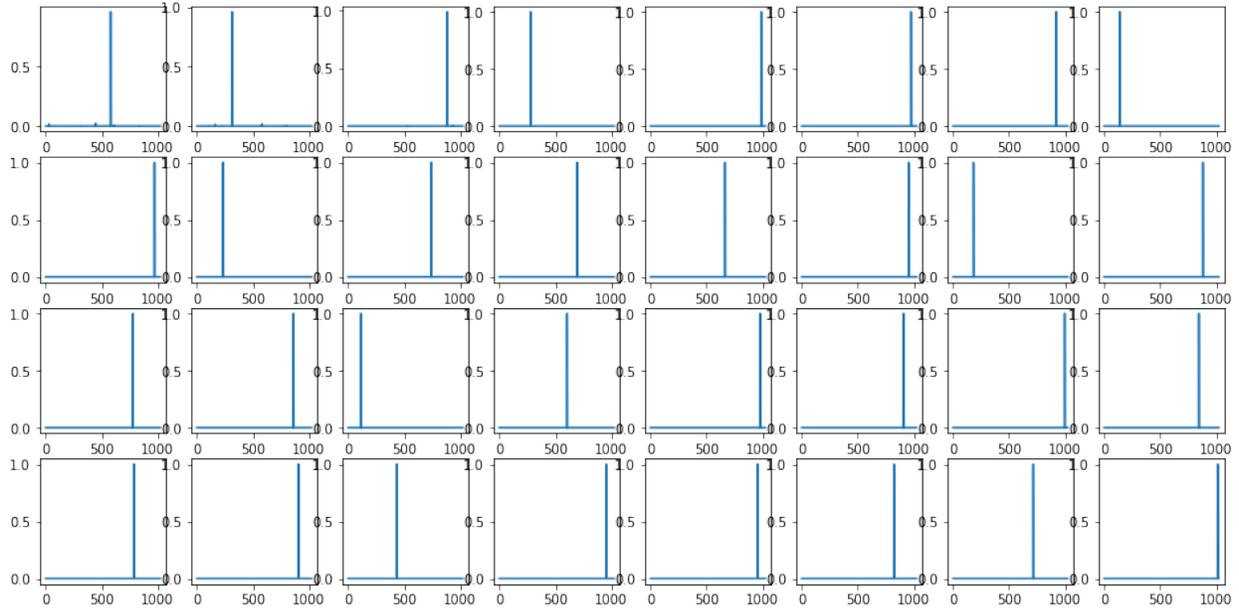

**Figure A5**. Conditional probability function of each of the 32 pixels in the bottom latent code that is used to sample the bottom latent code.

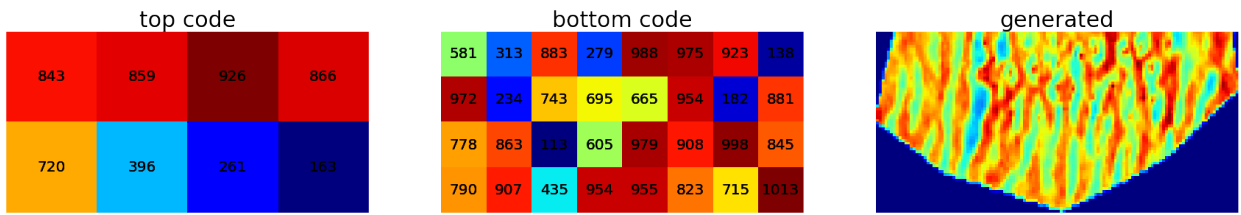

**Figure A6**. Newly sampled top and bottom latent codes obtained using the conditional probability functions shown in Figures A4 and A5. VQ-VAE-2 decoder uses the sampled top and bottom latent codes to generate a new geomodel, shown on the right.

**APPENDIX B: DEEP NEURAL NETWORK ARCHITECTURE OF PixelSNAIL**

Before passing a batch of images into the PixelBlock (shown in Figure B1), operations such as shifting and cropping is needed to ensure causality and resolve the blind spot problem. Gated residual block (GRB) is one of the building blocks for a deep PixelSNAIL network, it is formed by a shortcut connection, several causal convolution layers and a gating function [25]. The architecture of a GRB is presented in Figure B2. Similar to typical residual blocks, GRB allows the information flows more freely by shortcut connections where the output of a layer is fed to the output of another layer deeper in the block. The GRB block also processes conditioning input by adding the input into the network, as shown in Figure B2.

The attention block is the key element of a PixelSNAIL network in that it allows the network to model long-range dependencies and thus leading to better density estimation [22]. The attention block consists of three components: the *query*, the *key* and the *value*. To make it simple, we skipped the detailed explanation of each component, readers can refer original paper developed by Google [26]. By introducing attention blocks, the network achieves much large receptive field that almost encompasses all regions before current location.

In our specific case, the PixelBlock is repeated 4 times ($n = 4$), the gated residual block is repeated $m = 4$ and $r = 0$.

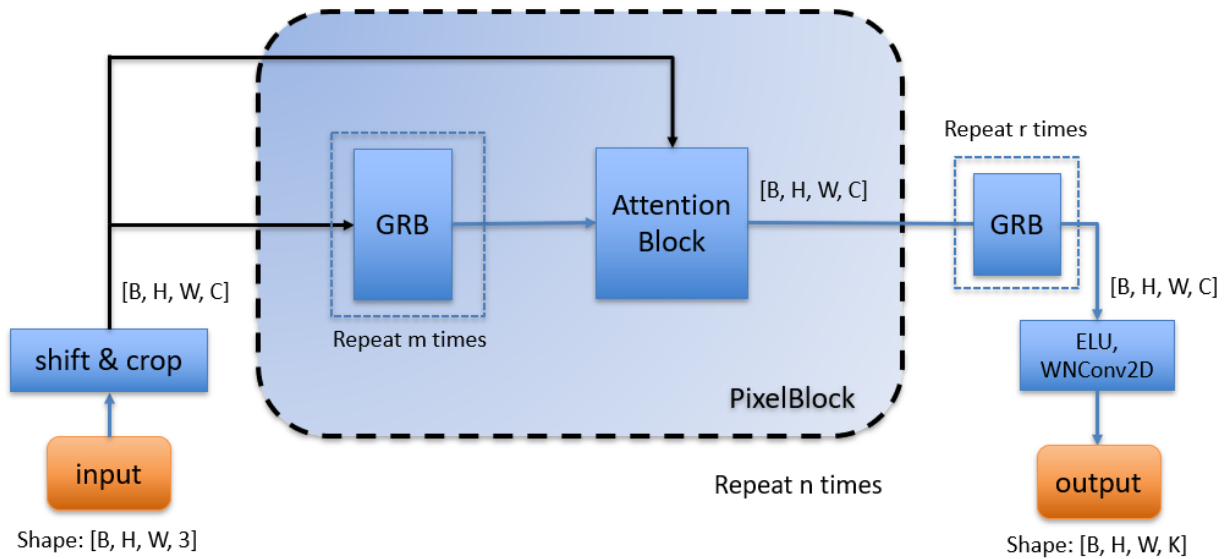

**Figure B1**. PixelSNAIL network architecture for building a self-supervised autoregressive model of the latent code distributions. In this figure, K: codebook size/number of classes

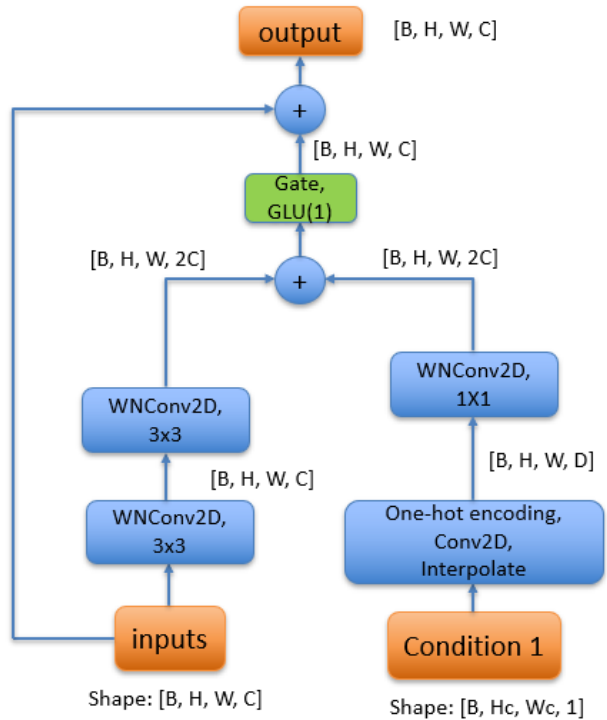

**Figure B2**. Gated residual block (GRB) architecture considering conditions, GRB is the building block for the entire PixelSNAIL model. It consists of several weight normalized convolution layers followed by a gate function. Also, a residual connection is added from the input to the output. The output dimension if equal to input dimension. In this figure, WNConv2D: 2D convolutional layer with weight normalization; Interpolate: up sampling the input to a given size.